\documentclass[a4paper]{article}
\usepackage{graphicx}
\usepackage{amsmath}

\begin{document}

\newcommand{\bin}[2]{\left(\begin{array}{c} \!\!#1\!\! \\  \!\!#2\!\! \end{array}\right)}
\newcommand{\threejm}[6]{\left(\begin{array}{ccc}#1 & #2 & #3 \\ #4 & #5 & #6 \end{array}\right)}

\huge

\begin{center}
The Hybrib Detailed / Statistical Opacity Code SCO-RCG: New Developments and Applications
\end{center}

\vspace{0.5cm}

\large

\begin{center}
Jean-Christophe Pain$^{a,}$\footnote{jean-christophe.pain@cea.fr (corresponding author)}, Franck Gilleron$^a$, Quentin Porcherot$^b$ and Thomas Blenski$^c$
\end{center}

\vspace{0.2cm}

\normalsize

\begin{center}
$^a$CEA, DAM, DIF, F-91297 Arpajon, France

$^b$DGA, 94110 Arcueil, France

$^c$CEA, DSM, IRAMIS, F-91191 Gif-sur-Yvette, France
\end{center}

\vspace{0.5cm}

\begin{abstract}
We present the hybrid opacity code SCO-RCG which combines statistical approaches with fine-structure calculations. Radial integrals needed for the computation of detailed transition arrays are calculated by the code SCO (Super-configuration Code for Opacity), which calculates atomic structure at finite temperature and density, taking into account plasma effects on the wave-functions. Levels and spectral lines are then computed by an adapted RCG routine of R. D. Cowan. SCO-RCG now includes the Partially Resolved Transition Array model, which allows one to replace a complex transition array by a small-scale detailed calculation preserving energy and variance of the genuine transition array and yielding improved high-order moments. An approximate method for studying the impact of strong magnetic field on opacity and emissivity was also recently implemented.
\end{abstract}

\section{DESCRIPTION OF THE CODE AND EFFECT OF DETAILED LINES}

When lines coalesce into broad unresolved patterns due to physical broadening mechanisms (Stark effect, auto-ionization, etc.), they can be handled by global methods \cite{Bauche88}. On the other hand, some transition arrays exhibit a small number of lines that must be taken into account individually. Those lines are important for plasma diagnostics, interpretation of spectroscopy experiments and for calculating the Rosseland mean, which is very sensitive to the gaps between lines in the spectrum. The hybrid opacity code SCO-RCG \cite{Porcherot11} combines statistical methods and fine-structure calculations, assuming local thermodynamic equilibrium. In order to decide whether a detailed treatment of lines is necessary or not and to determine the validity of statistical methods, the code uses criteria to quantify the porosity (localized absence of lines) of transition arrays. Data required for the calculation of lines (Slater, spin-orbit and dipolar integrals) are provided by SCO (Superconfiguration Code for Opacity) \cite{Blenski00}, which takes into account plasma screening and density effects on the wave-functions. Then, level energies and lines are calculated by an adapted routine (RCG) of Cowan's atomic-structure code \cite{Cowan81} performing the diagonalization of the Hamiltonian matrix. Transition arrays for which a DLA (Detailed Line Accounting) treatment is not required or impossible are described statistically, by UTA (Unresolved Transition Array), SOSA (Spin-Orbit Split Array) or STA (Super Transition Array) formalisms used in SCO. SCO-RCG calculations are restricted to a particular type of superconfiguration, in which all supershells are made of individual orbitals up to a limit beyond which all the remaining orbitals are gathered into a large final supershell, consistent with Inglis-Teller limit \cite{Inglis39}. The total opacity is the sum of photo-ionization, inverse Bremsstrahlung and Thomson scattering spectra calculated by SCO code and a photo-excitation spectrum arising from contributions of SCO and Cowan's codes (see Fig. 1) in the form

\begin{equation}
\kappa\left(h\nu\right)=\frac{1}{4\pi\epsilon_0}\frac{\mathcal{N}}{A}\frac{\pi e^2h}{mc}\sum_{X\rightarrow X'}f_{X\rightarrow X'}\mathcal{P}_X\Psi_{X\rightarrow X'}(h\nu),
\end{equation}

\noindent where $h$ is Planck's constant, $\mathcal{N}$ the Avogadro number, $\epsilon_0$ the vacuum polarizability, $m$ the electron mass, $A$ the atomic number and $c$ the speed of light. $\mathcal{P}$ is a probability, $f$ an oscillator strength, $\Psi(h\nu)$ a profile and the sum $X\rightarrow X'$ runs over lines, UTAs, SOSAs or STAs of all ion charge states present in the plasma. Special care is taken to calculate the probability of $X$ (which can be either a level $\alpha J$, a configuration $C$ or a superconfiguration $S$) because it can be the starting point for different transitions (DLA, UTA, SOSA, STA). In order to ensure the normalization of probabilities, we introduce three disjoint ensembles: $\mathcal{D}$ (detailed levels $\alpha J$), $\mathcal{C}$ (configurations $C$ too complex to be detailed) and $\mathcal{S}$ (superconfigurations $S$ that do not reduce to ordinary configurations). The total partition function then reads

\begin{equation}
U_{\mathrm{tot}}=U\left(\mathcal{D}\right)+U\left(\mathcal{C}\right)+U\left(\mathcal{S}\right)\;\;\;\;\mathrm{with}\;\;\;\;\mathcal{D}\cap\mathcal{C}\cap\mathcal{S}=\emptyset,
\end{equation}

where each term is a trace over quantum states of the form Tr$\left[e^{-\beta\left(\hat{H}-\mu\hat{N}\right)}\right]$, where $\hat{H}$ is the Hamiltonian, $\hat{N}$ is the number operator, $\mu$ the chemical potential and $\beta=1/\left(k_BT\right)$. The probabilities of the different species of the $N$-electron ion are

\begin{equation}\label{probs}
\begin{array}{ll}
\mathcal{P}_{\alpha J}=\frac{1}{U_{\mathrm{tot}}}\left(2J+1\right)e^{-\beta\left(E_{\alpha J}-\mu N\right)} & \mathrm{for~a~level},\\
\mathcal{P}_C=\frac{1}{U_{\mathrm{tot}}}g_C~e^{-\beta\left(E_C-\mu N\right)} & \mathrm{for~a~configuration},\\
\mathcal{P}_S=\frac{1}{U_{\mathrm{tot}}}\sum_{C\in S}g_C~e^{-\beta\left(E_C-\mu N\right)} & \mathrm{for~a~superconfiguration}.
\end{array}
\end{equation}

In SCO-RCG, configuration mixing is limited to electrostatic interaction between relativistic sub-configurations ($n\ell j$ orbitals) belonging to a non-relativistic configuration ($n\ell$ orbitals). In order to complement DLA (Detailed Line Accounting) efforts, the code was recently improved \cite{Pain13} with the PRTA (Partially Resolved Transition Array) model \cite{Iglesias12}, which may replace the single feature of a UTA by a small-scale detailed transition array that conserves the known transition-array properties (energy and variance) and yields improved higher-order moments. In the PRTA approach, open subshells are split in two groups. The main group includes the active electrons and those electrons that couple strongly with the active ones. The other subshells are relegated to the secondary group. A small-scale DLA calculation is performed for the main group (assuming therefore that the subshells in the secondary group are closed) and a statistical approach for the secondary group assigns the missing UTA variance to the lines. In the case where the transition $C\rightarrow C'$ is a UTA that can be replaced by a PRTA (see Fig. 2), its contribution to the opacity is modified according to

\begin{equation}
f_{C\rightarrow C'}~\mathcal{P}_C~\Psi_{C\rightarrow C'}(h\nu)\approx\sum_{\bar{\alpha}\bar{J}\rightarrow\bar{\alpha'}\bar{J'}}f_{\bar{\alpha}\bar{J}\rightarrow\bar{\alpha'}\bar{J'}}~\mathcal{P}_{\bar{\alpha}\bar{J}}~\Psi_{\bar{\alpha}\bar{J}\rightarrow\bar{\alpha'}\bar{J'}}(h\nu),
\end{equation}

where the sum runs over PRTA lines $\bar{\alpha}\bar{J}\rightarrow\bar{\alpha'}\bar{J'}$ between pseudo-levels of the reduced configurations, $f_{\bar{\alpha}\bar{J}\rightarrow\bar{\alpha'}\bar{J'}}$ is the corresponding oscillator strength and $\Psi_{\bar{\alpha}\bar{J}\rightarrow\bar{\alpha'}\bar{J'}}$ is the line profile augmented with the statistical width due to the other (non included) spectator subshells. The probability of the pseudo-level $\bar{\alpha}\bar{J}$ of configuration $\bar{C}$ reads

\begin{equation}
\mathcal{P}_{\bar{\alpha}\bar{J}}=\frac{\left(2\bar{J}+1\right)e^{-\beta\left(E_{\bar{\alpha}\bar{J}}-\mu N\right)}}{\sum_{\bar{\alpha}\bar{J}\in\bar{C}}\left(2\bar{J}+1\right)e^{-\beta \left(E_{\bar{\alpha}\bar{J}}-\mu N\right)}}\times\mathcal{P}_C\;\;\;\;\mathrm{with}\;\;\;\;\sum_{\bar{\alpha}\bar{J}\in\bar{C}}\mathcal{P}_{\bar{\alpha}\bar{J}}=\mathcal{P}_C,
\end{equation}

where $\mathcal{P}_C$ is the probability of the genuine configuration given in Eq. (\ref{probs}).
 
\begin{figure}
\begin{center}
\includegraphics[width=10cm]{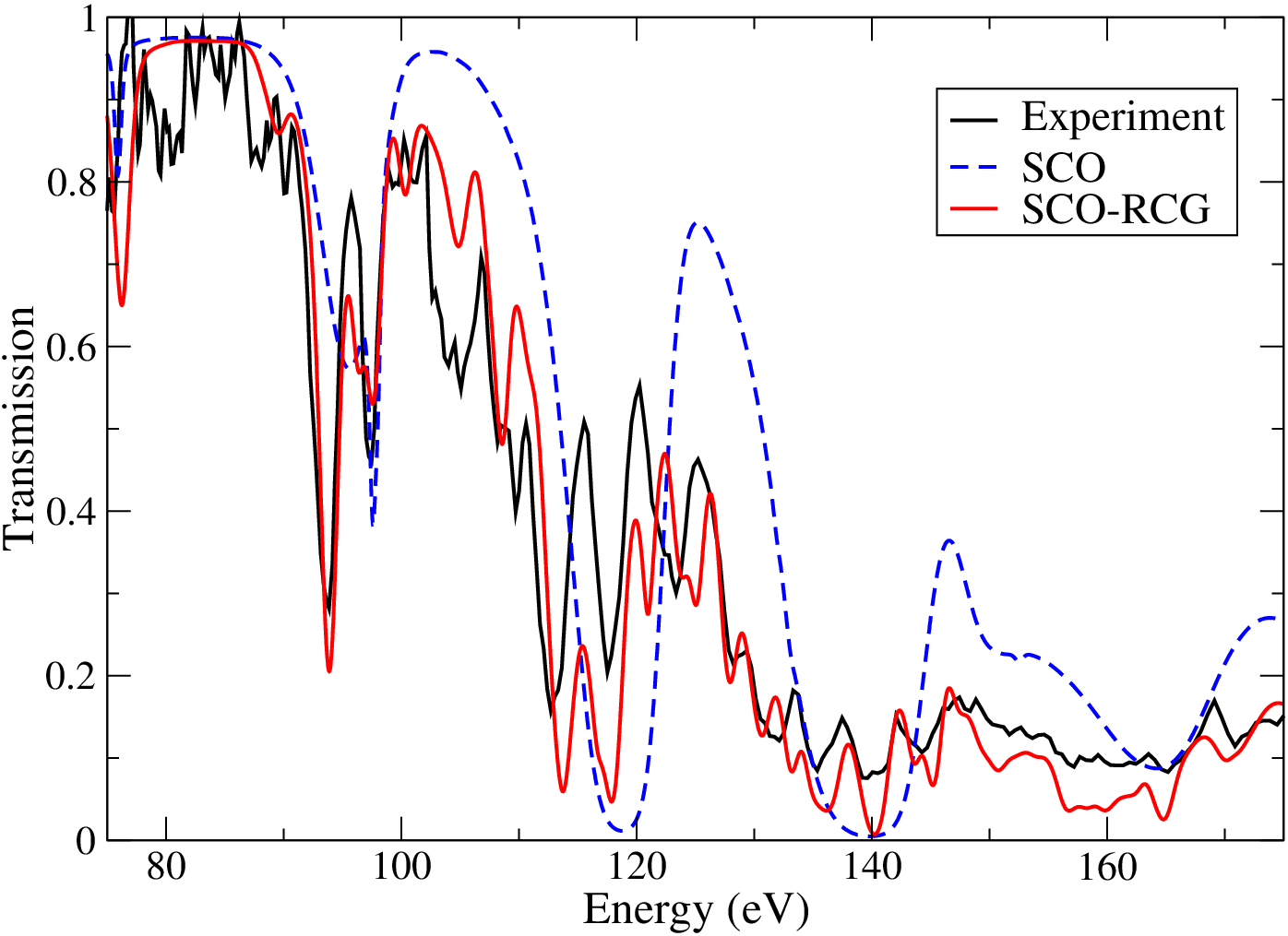}
\caption{(Color online) Interpretation with SCO-RCG code of the aluminum spectrum ($2p\rightarrow 3s$ transitions) measured by Winhart et al. \cite{Eidmann98}. Temperature gradients are simulated by an average over four temperatures: 18, 20, 22 and 24 eV. The full statistical SCO calculation at $T$=20 eV is also represented.}\label{fig:1}
\end{center}
\end{figure}

\begin{figure}
\begin{center}
\includegraphics[width=10cm]{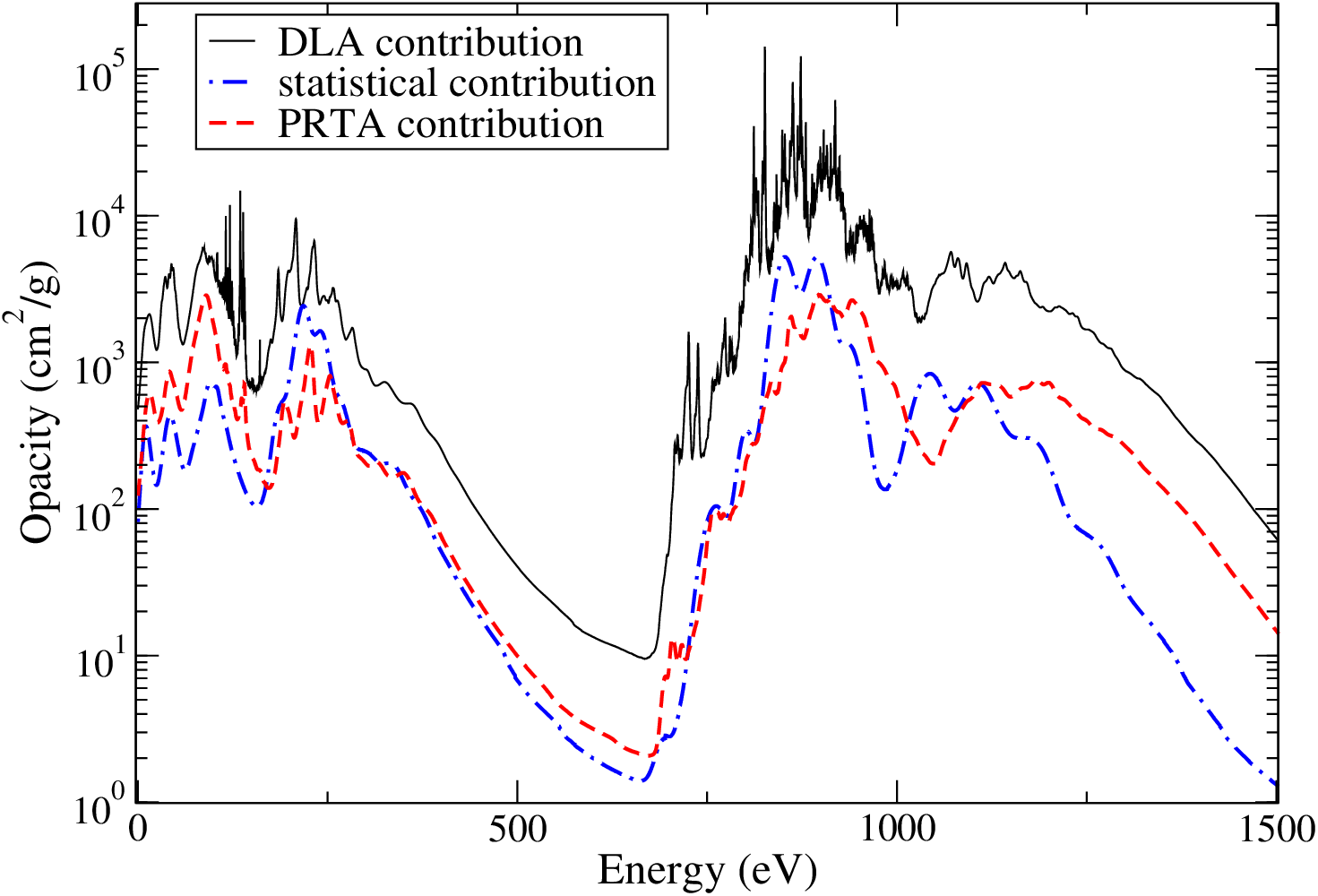}
\caption{(Color online) The three independent contributions to photo-excitation calculated by SCO-RCG code for an iron plasma at $T$=192.91 eV and $\rho$=0.578 g.cm$^{-3}$ (boundary of the convective zone of the Sun).}\label{fig:2}
\end{center}
\end{figure}

\begin{figure}
\begin{center}
\includegraphics[width=10cm]{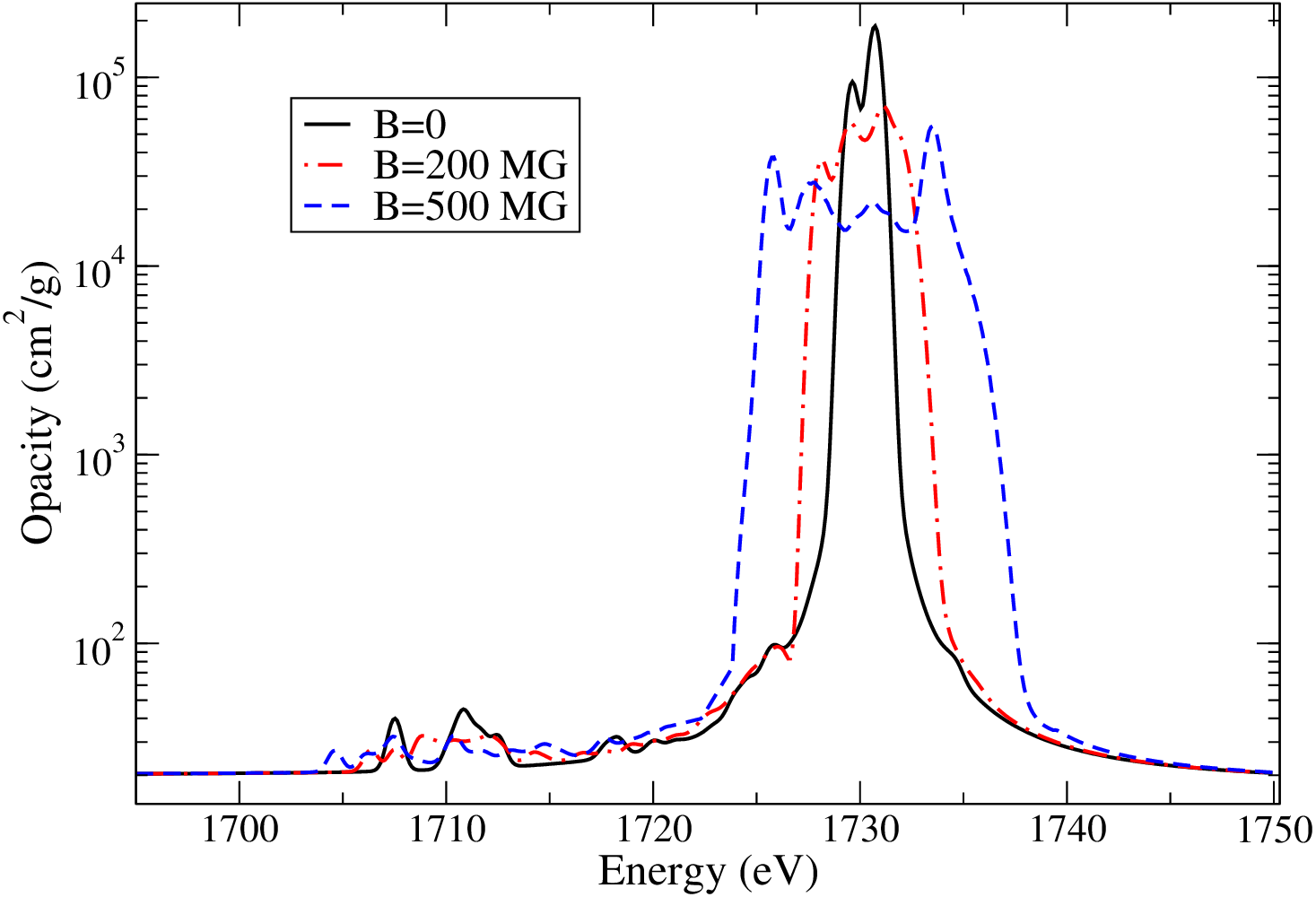}
\caption{(Color online) SCO-RCG calculations with and without magnetic field for an aluminum plasma at $T$=1 keV and $\rho$=2.~10$^{-2}$ g.cm$^{-3}$ (transitions $1s\rightarrow 2p$).}\label{fig:3}
\end{center}
\end{figure}

\begin{table}
\begin{center}
\begin{tabular}{|c|c|c|c|}\hline
\multicolumn{2}{|c|}{} & $J_2=J_1$ & $J_2=J_1\pm 1$ \\\hline
$\sigma_q$ & $\alpha_3$ &  \multicolumn{2}{c|}{$(-1)^qq\left(J_1-J_2\right)\mathrm{sgn}\left[g_1-g_2\right]\frac{2\sqrt{5}}{3\sqrt{3}}\frac{J_>}{\sqrt{J_<\left(J_>+1\right)}}$} \\\cline{2-4}
& $\alpha_4$ & $\frac{5}{7}\left(\frac{12J_1\left(J_1+1\right)-17}{4J_1\left(J_1+1\right)-3}\right)$ & $\frac{5}{21}\left(13-\frac{4}{J_<\left(J_>+1\right)}\right)$ \\ 
\hline
$\pi$ & $\alpha_3$ & \multicolumn{2}{c|}{0} \\\cline{2-4}
& $\alpha_4$ & $\frac{25}{7}\left(\frac{3\left[\left(J_1+2\right)J_1^2-1\right]J_1+1}{\left[1-3J_1\left(J_1+1\right)\right]^2}\right)$ & $\frac{5}{7}\left(3-\frac{2}{J_<\left(J_>+1\right)}\right)$ \\\hline
\end{tabular}
\caption{Values of $\alpha_3$ and $\alpha_4$ of the Zeeman components. $J_<=\min(J_1,J_2)$ and $J_>=\max(J_1,J_2)$. $\mathrm{sgn}\left[x\right]$ is the sign of $x$.}
\label{tab:a}
\end{center}
\end{table}

The SCO-RCG code has been successfully compared to experimental spectra. The comparisons in Fig. \ref{fig:1} shows the relevance of the hybrid model and the limits of a full statistical calculation. Among others, the code is also used for astrophysical applications \cite{Gilles11,Turck11}. Figure \ref{fig:2} represents the different contributions to opacity (DLA, statistical and PRTA) for an iron plasma in conditions corresponding to the boundary of the convective zone of the Sun.

\section{STATISTICAL MODELING OF ZEEMAN EFFECT}

\noindent Quantifying the impact of a magnetic field on spectral line shapes is important in astrophysics, in inertial confinement fusion or for Z-pinch experiments. Because the line computation becomes even more tedious in that case, we propose, in order to avoid the diagonalization of the Zeeman hamiltonian, to describe Zeeman patterns in a statistical way. In the presence of a magnetic field $B$, a level $\alpha J_1$ (energy $E_1$) splits into $2J_1+1$ states $M_1$ ($-J_1\leq M_1\leq J_1$) of energy $E_1+\mu_Bg_1M_1$ , $\mu_B$ being the Bohr magneton and $g_1$ the Land\'e factor in intermediate coupling (provided by RCG routine). Each line splits in three components associated to selection rule $\Delta M$=$q$, where $q$=0 for a $\pi$ component and $\pm 1$ for a $\sigma_{\pm}$ component. The intensity of a component can be characterized by the strength-weighted moments of the energy distribution. The $n^{th}-$ order moment reads

\begin{equation}
\mathcal{M}_n\left[q\right]=3\sum_{M_1,M_2}\threejm{J_1}{1}{J_2}{-M_1}{-q}{M_2}^2\left(E_2-E_1+\mu_BB\left[g_2M_2-g_1M_1\right]\right)^n,
\end{equation}
which can be evaluated analytically \cite{Pain12a,Pain12b}, using graphical representation of Racah algebra or Bernoulli polynomials \cite{Mathys87}. A good representation of the Zeeman profile is obtained using, for each component, the fourth-order Gram-Charlier expansion series:

\begin{equation}
\Psi_Z(u)=\frac{1}{\sqrt{2\pi v}}\exp\left(-\frac{u^2}{2}\right)\left[1-\frac{\alpha_3}{2}\left(u-\frac{u^3}{3}\right)+\frac{\left(\alpha_4-3\right)}{24}\left(3-6u^2+u^4\right)\right],
\end{equation}
where $u=\left(E-\mathcal{M}_1\right)/\sqrt{v}$, $v=\mathcal{M}_2-\left(\mathcal{M}_1\right)^2$ is the variance and the reduced centered moments are defined as
 
\begin{equation}
\alpha_n=\left(\sum_{k=0}^n\bin{n}{k}\mathcal{M}_k\left(-\mathcal{M}_1\right)^{n-k}\right)/v^{n/2},
\end{equation}

where $\alpha_3$ (skewness) and $\alpha_4$ (kurtosis) quantify respectively the asymmetry and the sharpness of the component (see Table \ref{tab:a}). The total line profile results from the convolution of $\Psi_Z$ with the other broadening mechanisms (see Fig. \ref{fig:3}). The contribution of a magnetic field to an UTA can be taken into account roughly by adding a contribution $2/3\;(\mu_BB)^2\approx 3.35\; 10^{-5}$ [$B$(MG)]$^2$ eV$^2$ to the statistical variance. The approximate method provides quite a good description of the effect of a strong magnetic field on spectral lines.

\section{CONCLUSION}

By combining different degrees of approximation of the atomic structure (levels, configurations and superconfigurations), the SCO-RCG code allows one to explore a wide range of applications, such as the calculation of Rosseland means, the generation of opacity tables, or the spectroscopic interpretation of high-resolution spectra. The PRTA model was recently adapted to the hybrid statistical / detailed approach in order to reduce the statistical part and fasten the calculations. An approximate approach providing a fast and quite accurate estimate of the effect of an intense magnetic field on opacity was also implemented. The formalism requires the moments of the Zeeman components of a line, which can be obtained analytically in terms of the quantum numbers and Land\'{e} factors. It was found that the fourth-order A-type Gram-Charlier expansion series provides better results than the usual development in powers of the magnetic field often used in radiative-transfer models. In the future, we plan to improve the treatment of Stark broadening in order to increase the capability of the code as concerns K-shell spectroscopy.

\end{document}